\documentclass[a4paper,10pt]{article}
\usepackage{amsfonts,amsmath,graphicx,a4wide}

\numberwithin{equation}{section}

\def\d{{\rm d}}

\begin{document}
\thispagestyle{empty}
\title{\Large{\bf Thin static charged dust Majumdar--Papapetrou shells with
high symmetry in $D~\geq~4.$}}

\author{{\sc Martin \v{C}erm\'{a}k}$^{a,b}$ and {\sc Martin~Zouhar}$^{c,b}$
\\~\\
$^{a}$Institute of Physics of Materials, \\\v{Z}i\v{z}kova 22, CZ 616 62 Brno, Czech Republic\\~\\
$^{b}$Department of Physical Electronics, \\Faculty of Science, Masaryk University Brno, \\Kotl\'{a}\v{r}sk\'{a} 2, Brno, Czech Republic\\~\\
$^{c}$CEITEC - Central European Institute of Technology,\\ c/o Masaryk University Brno,\\ \v{Z}erot\'{i}novo n\'{a}m. 9, CZ 601 77 Brno, Czech Republic}
\maketitle
\vfill
\begin{abstract}
We present a systematical study of static $D~\geq~4$ space--times of high symmetry with the matter source being a thin charged dust hypersurface shell.
The shell manifold is assumed to have the following structure $\mathbb{S}_{\beta}\times\mathbb{R}^{D-2-\beta},$ $\beta\in\{0,\ldots,D-2\}$ is dimension of a sphere $\mathbb{S}_{\beta}.$
In case of $\beta~=~0,$ we assume that there are two parallel hyper--plane shells instead of only one.

The space--time has Majumdar--Papapetrou form and it inherits the symmetries of the shell manifold -- it is invariant under both rotations of the $\mathbb{S}_{\beta}$ and translations along $\mathbb{R}^{D-2-\beta}.$

We find a general solution to the Einstein--Maxwell equations with a given shell.
Then, we examine some flat interior solutions with special attention paid to $D~=~4.$
A connection to $D~=~4$ non--relativistic theory is pointed out.
We also comment on a straightforward generalisation to the case of Kastor--Traschen space--time, i.e. adding a non--negative cosmological constant to the charged dust matter source.
\section*{Keywords}Majumdar--Papapetrou \and Kastor--Traschen \and higher dimensional \and thin shell \and charged dust \and general relativity.
\end{abstract}

\section{Introduction}
\label{intro}
Finding exact solution to Einstein equations is a
difficult task in general. The equations of motion simplify if the
sought solution has a symmetry, therefore solutions of high
symmetry are relatively easy to study and hence quite well known.

\medskip
Some modern physical models and theories abandon the assumption
that the space--time we live in has four dimensions. The extension
to higher dimensions can be motivated by unification attempts,
such as Kaluza--Klein (unification of gravity and gauge
fields)~\cite{kaluza.klein} and string theory (candidate for a
quantum gravity theory), or brane--world model, where our world is
a brane embedded in a higher dimensional total space--time called
bulk (which may explain why gravity, "leaking" to the bulk, is
weaker than other interactions that are trapped on the
brane)~\cite{brane.scale}.

The above mentioned models and theories give a physical motivation
to examine higher dimensional problems that can be interesting
mathematical problem by itself.

\medskip
Let us present a reason to study shells. Potential (both
electrostatic and gravitational) of point like and line like
sources diverges at the source, probably due to "too singular"
nature of the source density. To avoid the singular behavior at
the point and line matter sources, it is possible to modify the
equations for the field, e.g. consider non--linear
electrodynamics~\cite{non.linear.EM.singularity} that can remove
singularities in the field strength tensor $F_{\alpha\beta},$ or
modify the matter source distribution, e.g. by considering shells.
We shall see that the potential at the shells considered in this
work has a finite value.

\smallskip
In $D~=~4$ space--time, some examples of the highly symmetric
solutions are static spherical, cylindrical and plane spatial
symmetry solutions to the Einstein-Maxwell equations, reviewed
e.g. in~\cite{stephani}. Particular examples of such solutions are
those of thin charged massive shells studied e.g.
in~\cite{metin,metin2}.

In case of a higher dimensional space-time, the above mentioned
three spatial symmetries can be generaliced in more ways, hence
the considered shell manifold
$\mathcal{M}~=~\mathbb{S}_{\beta}\times\mathbb{R}^{D-2-\beta},$
see~\eqref{the.D.shells}.

\medskip
The Majumdar--Papapetrou form metric has already been studied in~\cite{gurses.reference,lemos.reference} among others.
When compared to~\cite{gurses.reference}, that examines a cloud of charged dust, we study a general dimension case and charged dust shells only and we present exact solutions in all the cases of the symmetries, classified by the value of parameter $\beta,$ considered.
This distinguishes our article also from reference~\cite{metin} the author of which extends his work in~\cite{gurses.reference} by adding rather general $D~=~4$ shells considerations with only one exact shell solution in the case of spherical symmetry.

The work~\cite{lemos.reference} comments on some general properties of a higher dimensional space--time of Majumdar--Papapetrou type with proper derivation of the metric form from a quite general setting.
A reader curious about the origin of metric in equation~\eqref{5,1} is referred there though he or she will find no exact solution there.

Moreover, the generalisation to Kastor--Traschen type space--time, missing in~\cite{metin,gurses.reference,lemos.reference}, is discussed in our article in $D~\geq~4$ which is, in some way, an extension of the $D~=~4$ treatment in~\cite{metin2}.

\bigskip
The shell separates the space-time into two regions that may be
called exterior region and interior region~\footnote{This
terminology is a bit cumbersome if the shell is not a boundary to
a compact volume.}. The two regions have to be matched across the
shell. The metric is continuous across the surface but a quantity
related to its first derivative need not be. In general, the first
derivative exhibits a jump across the shell. The jump is related
to the shell matter source.

We integrate the equations of motion allowing distributions (such
as Dirac and Heaviside "functions") enter the expressions for
metric components. This allows us to write each exterior and
interior solution component in a single formula.

In a special case, the interior space--time region is flat and the
exterior is a curved solution of Einstein-Maxwell equations.

If we demand the solution to be non-collapsing, then we have to
enforce that repulsive electrostatic forces are in balance with
the attractive gravity forces.

Thus obtained balance condition will be examined in both general
relativity and non-relativistic physics in this article.

In this work, we consider standard Einstein--Hilbert general
relativity with Levi--Civita connection, the signature is mostly
minus, see e.g.~\cite{horsk,landau}.

Greek indices (run from $0$ to $D~-~1$) are used for space-time
coordinates and Latin indices (run from $1$ to $D~-~1$) will be
used for spatial coordinates. $x^0~=~ct.$ Partial derivative is
denoted by comma, covariant derivative by semicolon.

We shall explicitly write out parameters of the solutions,
including universal constants (light velocity $c,$ Newton
gravitational constant $\kappa$ and permitivity of vacuum
$\varepsilon_0$).

Outline of this article is: We give some general comments on the
equations of motion in the case of matter sources of interest in
Section~\ref{eom.general}. The solutions have the
Majumdar--Papapetrou metric form discussed in
Section~\ref{mp.metric}. We solve the Einstein--Maxwell equations
for the Majumdar--Papapetrou metric for the shells considered in
Section~\ref{shells.general.D}. Consequently, some sub--cases
corresponding to $D\in\{4,5\},$ and subject to additional
conditions, are discussed in detail in Section~\ref{shells.D.4.5}.
Then, the non--relativistic approximation follows in
Section~\ref{non.relativistic}. Consequently, we briefly discuss
generalisation to Kastor--Traschen space--time with positive
cosmological constant in Section~\ref{Kastor.Traschen.lambda}.

\section{General notes on equations of motion}\label{eom.general}
The field equations for metric, the Einstein equations, are
\begin{equation}
R_{\alpha\beta}-\frac{1}{2}g_{\alpha\beta}R\equiv
G_{\alpha\beta}=KT_{\alpha\beta} ,\label{1}\end{equation} where
$K$ is coupling constant. $K~=~\frac{8\pi \kappa }{c^{4}}$ in a
four dimensional space-time.


The total stress-energy-momentum tensor of a charged dust consist
of two parts - the dust and the electro-magnetic field
contributions.
\begin{equation}
T_{\alpha\beta}=\rho_{m}c^{2}u_{\alpha}u_{\beta}\frac{\d s}{\d
x^{0}}
+\varepsilon_{0}c^{2}\left[\frac{1}{4}F^{\gamma\delta}F_{\gamma\delta}g_{\alpha\beta}
-F_{\alpha}\;^{\delta}F_{\beta\delta}\right].
\label{3}\end{equation} First part represents the dust and the
second represents the electro-magnetic field. The field strength
tensor $F_{\alpha\beta}$ is related to the gauge potential
$A_{\alpha}$ as $F_{\mu\nu}~=~A_{\nu;\mu}~-~A_{\mu;\nu}.$

Since we seek for a static solution, we pick up a coordinate
system in which the dust is at rest. As a result, velocity has
only one non--zero component so that $u_{\alpha} =
\sqrt{g_{00}}\delta^0_{\alpha}$ and derivative of the line element
$s$ by $x^{0}$ is
$\frac{\mathrm{d}s}{\mathrm{d}x^{0}}=\sqrt{g_{00}}.$

Volume mass density of the shells is mass surface density
multiplied by the delta function
$\rho_{\textup{m}}=\mu\delta(r-r_{0})$, where $\mu$ is constant mass
density on the shell, $\delta(r-r_{0})$ is the delta function, $r$ is
coordinate that is constant on the shell.

The only non-zero component of the dust (i.e. massive part of the
total) stress-energy-momentum tensor is
\begin{equation}
T_{\textup{m}00}=\mu
c^{2}g_{00}^{\frac{3}{2}}\delta(r-r_{0}).\label{3,1}
\end{equation}
Volume charge density is
$\rho_{\textup{e}}=\frac{\sigma\delta(r-r_{0})}{\sqrt{-g_{rr}}},$
where $\sigma$ is surface charge density and $g_{rr}$ is $rr$
component of the metric tensor. We shall use adapted coordinate
system in which $r~=~x^1$.

The equation of motion for the electromagnetic field relates the
strength tensor $F_{\alpha\beta}$ to electric current $j^{\alpha}$
as
\begin{equation}
{F}^{\alpha\beta}_{\;\;\;\; ;\alpha} = \frac{1}{\sqrt{|g|}}\left(
\sqrt{|g|}{F}^{\alpha\beta}\right)_{,\alpha} =
\frac{j_{\textup{e}}^{\beta}}{\varepsilon_{0}c^{2}},\label{4}
\end{equation}
where we have used identity relating covariant divergence of an
anti-symmetric tensor with ordinary divergence.

If the source is at rest, then the current has only one non--zero
component given by
$j_{\textup{e}}^{0}=\frac{c\rho_{\textup{e}}}{\sqrt{g_{00}}}.$
Charge can be calculated using
$\textup{d}Q=\rho_{\textup{e}}\sqrt{\frac{|g|}{g_{00}}}\textup{d}V$,
where $\textup{d}V~=~\prod^{D-1}_{i=1}\d x^i$ and $g$ is
determinant of the metric tensor.

Mass can be calculated in a similar spirit as
$\textup{d}m=\rho_{\textup{m}}\sqrt{|g|}\textup{d}V$.

Using the expression for the charge current components, the
equation~\eqref{4} becomes
\begin{equation}
\left(\sqrt{|g|}{F}^{\alpha0}\right)_{,\alpha}=\frac{\sqrt{|g|}}{\sqrt{g_{00}}}\frac{\rho_{\textup{e}}}{\varepsilon_{0}c},\;
\left(\sqrt{|g|}{F}^{\alpha i}\right)_{,\alpha}=0
.\label{5}\end{equation}

%

\section{Einstein equations for Majumdar--Papapetrou metric}\label{mp.metric}
Majumdar--Papapetrou metric describes space-time of charged matter
with attractive gravitational force balanced by repulsive
electrostatic force. The Majumdar--Papapetrou metric can also
describe static charged dust shells.

A general form of the metric tensor is given by~\cite{lemos.reference,mitra}
\begin{equation}
\mathrm{d} s^2 = G^{-2}(c\,\mathrm{d}t)^2 -
G^{2/(D-3)}\,\gamma_{ij}\d x^i\d x^j ,\label{5,1}\end{equation}
where $\gamma_{ij}$ is a flat $(D~-~1)$-dimensional metric, i.e.
$$\gamma_{ij}\d x^i\d x^j = \delta_{kl}\d \tilde{x}^k\d \tilde{x}^l = (\d\vec{\tilde{x}})^{2},$$
where $\tilde{x}^k$ are Cartesian coordinates in the above
formula.

Consider static Majumdar--Papapetrou $D-$dimensional metric in Cartesian coordinate system.
We drop the tilde over $x$ in 
the following expressions.

The corresponding Einstein tensor has following non--zero components
\begin{equation}
G^{\rm MP}_{00} =\frac{1}{2}\frac{D-2}{D-3} \frac{(\nabla G)^2 -
2G\Delta G}{G^{4 + 2/(D-3)}},\; G^{\rm MP}_{ij} =
\frac{1}{2}\frac{D-2}{D-3}\frac{(\nabla G)^2\delta_{ij} -
2G_{,i}G_{,j}}{G^2} ,\label{5,2}\end{equation} where we have
defined quantities related to flat space operators -- square of a
scalar gradient and Laplace operator -- as
\begin{equation}
(\nabla G)^2 = \gamma^{ij}G_{,i}G_{,j},\; \Delta G =
\frac{1}{\sqrt{|\gamma|}}\left(\sqrt{|\gamma|}\gamma^{ij}G_{,i}\right)_{,j},\;
|\gamma|\equiv\left|{\rm det}\,(\gamma_{ij})\right|
.\label{Laplace.flat.general}\end{equation} The electrostatic
field is given by
\begin{equation}
A_{\alpha}=\frac{\phi_{\textup{e}}(x^i)G^{-2}}{c}\delta^0_{\alpha},\;
F_{\alpha\beta}=
\delta^0_{\beta}A_{0,\alpha} - \delta^0_{\alpha}A_{0,\beta},\;
F^{i0} =
-\delta^{ij}\frac{(\phi_{\textup{e}}G^{-2})_{,j}}{c}G^{2\frac{D-4}{D-3}}
,\label{electro.static}\end{equation} where
$\phi_{\textup{e}}~\equiv~cA^0$ is electrostatic potential.

The non--zero components of the EM metric stress-energy-momentum
tensor are
written in terms of $A_0$ -- instead of the function
$\phi_{\textup{e}}$ -- for the sake of brevity
\begin{equation}
T_{\textup{e}00} = \frac{\varepsilon_{0}c^2}{2} \frac{(\nabla
A_0)^2}{G^{2/(D-3)}},\; T_{\textup{e}ij} =
\frac{\varepsilon_{0}c^2}{2}\left[ (\nabla A_0)^2\delta_{ij} -
2A_{0,i}A_{0,j}\right]G^2 ,\label{5,4}\end{equation} where
$(\nabla A_0)^2~=~\gamma^{ij}A_{0,i}A_{0,j}$ was introduced in the
same way as $(\nabla G)^2.$

It is easy to show, using~\eqref{3,1}, that the only non-zero
component of the dust stress-energy-momentum tensor becomes
\begin{equation}
T_{\textup{m}00}=\frac{c^{2}\rho_{\textup{m}}}{G^{3}}.\label{5,5}
\end{equation}

Einstein equations relate the metric function $G,$ electrostatic
potential $\phi_{\textup{e}}$ and density of mass
$\rho_{\textup{m}}$ in the following way~\cite{hawking}
\begin{equation}
\frac{1}{G}=C\mp\sqrt{K\varepsilon_{0}}\sqrt{\frac{D-3}{D-2}}\frac{\phi_{\textup{e}}}{G^{2}},
\;\Delta
G~=~-~\frac{D-3}{D-2}Kc^{2}G^{\frac{2}{D-3}}\rho_{\textup{m}}
,\label{5,51}\end{equation} where $\Delta$ is the flat space
Laplace operator, $C$ is a constant of integration and $K$ is
coupling constant. The Einstein equation have so far allowed for
an ambiguity in sign relating $G$ and $\phi_{\textup{e}},$ hence
the $\mp.$

The mathematical solution for $\phi_{\textup{e}}$ in terms of the
metric function $G$ presented in~\eqref{5,51} contains an
integration constant $C$ which we shall put equal zero in order to
obtain a good correspondence with non--relativistic physics, then
$G=\mp\sqrt{K\varepsilon_{0}}\sqrt{\frac{D-3}{D-2}}\phi_{\textup{e}}$.

The EM field equations of motion~\eqref{5}, together with
both~\eqref{electro.static} and~\eqref{5,51}, lead to
\begin{eqnarray*}
G^{\frac{D-1}{D-3}}\frac{\rho_{\textup{e}}}{\varepsilon_{0}c} &=&
\frac{\sqrt{|g|}}{\sqrt{g_{00}}}\frac{\rho_{\textup{e}}}{\varepsilon_{0}c}
=
\left(\sqrt{|g|}F^{\alpha0}\right)_{,\alpha}=\left(\sqrt{|g|}F^{i0}\right)_{,i}
= \left(\sqrt{|g|}F_{j0}g^{00}g^{ji}\right)_{,i}
\\
&=&
\left[-\sqrt{|g|}g^{00}g^{ji}\frac{(\phi_{\textup{e}}G^{-2})_{,j}}{c}\right]_{,i}
=
\left[-G^2\delta^{ij}\frac{(\phi_{\textup{e}}G^{-2})_{,j}}{c}\right]_{,i}
\\
&=&
\mp\sqrt{\frac{D-2}{D-3}}\frac{1}{c\sqrt{K\varepsilon_0}}\delta^{ij}G_{,ji}
= \mp\sqrt{\frac{D-2}{D-3}}\frac{1}{c\sqrt{K\varepsilon_0}}\Delta
G
\\
&=& \pm
c\sqrt{\frac{D-3}{D-2}}\sqrt{\frac{K}{\varepsilon_0}}G^{\frac{2}{D-3}}\rho_{\textup{m}}
.\end{eqnarray*}

Thus it implies balance condition for mass and charge densities
\begin{equation}
c^{2}\sqrt{K\varepsilon_{0}\frac{D-3}{D-2}}\rho_{\textup{m}}=
|\rho_{\textup{e}}|G .\label{5,6}\end{equation} We have written an
absolute value of charge density to ensure mass density is
positive. This corresponds to a certain choice of the $\pm$
factor, i.e. the sign ambiguity is removed.

The condition~\eqref{5,6} can be written using the total
quantities by using integral relations
\begin{equation}
M=\int_{V}\sqrt{|g|}\rho_{\textup{m}}\mathrm{d}V,\;
Q=\int_{V}\sqrt{\frac{|g|}{g_{00}}}\rho_{\textup{e}}\mathrm{d}V
,\label{5,61}\end{equation} where the integration domain $V$ must
contain the shell, i.e. $\mathcal{M}~\subset~V.$

The balance condition becomes
\begin{equation}
Mc^{2}\sqrt{K\varepsilon_0\frac{D-3}{D-2}}=|Q|.\label{5,7}
\end{equation}
We put
$K=\frac{D-2}{D-3}\frac{4\pi\kappa}{c^{4}}$~\footnote{\label{coupling}
Two approaches to the coupling constants can be found in the
literature. The first assumes that the Poisson equation, obtained
in non--relativistic limit, retains the same form in higher
dimensions. The second corrects the coupling constant in the
Poisson equation so that it explicitly depends on the space--time
dimension~\cite{merb}. }. Then the balance condition~\eqref{5,7}
is independent of number of space--time dimensions
\begin{equation}
M\sqrt{4\pi\kappa\varepsilon_{0}}=|Q| .\label{5,8}\end{equation}


\bigskip
We have found general form of the solution -- the electrostatic
potential $\phi_{\textup{e}}$ as a function of the metric function
$G,$ equation~\eqref{5,51}, a formula relating mass and charge
densities of a charged dust, equation~\eqref{5,6}. Moreover, the
equation~\eqref{5,51} also includes the relation between $G$ and
mass density $\rho_{\textup{m}}.$

Thus it suffices to prescribe e.g. the mass density
$\rho_{\textup{m}},$ solve the second relation in
equation~\eqref{5,51} with respect to $G$ and the space--time is
fully determined. This will be done in the next section.

\section{Thin shell solution of high symmetry in $D\geq 4$}\label{shells.general.D}
We shall consider charged dust shells being hypersurfaces
$\mathcal{M}$ in the Euclidean space $(\mathbb{R}^{D-1},\gamma).$
\begin{equation}
\mathbb{R}^{D-1}\supset \mathcal{M} =
\mathbb{S}_{\beta}\times\mathbb{R}^{D-2-\beta},\; \dim\mathcal{M}
= D-2,\;
\beta\in\{0,1,\ldots, D-2\} .\label{the.D.shells}\end{equation}

Since the shells considered are of high symmetry, we shall work in
a coordinate system adapted to the symmetries instead of the
Cartesian one used in the previous section. We are going to denote
$r~=~x^1$ a coordinate describing distance from the shell in such
an adapted coordinate system.

We may express it as a function of spatial Cartesian coordinates
$x^{i}$ in the following way $r~=~\sqrt{\sum_{i=1}^{\beta+1}(x^{i})^{2}},$ where
$\beta$ is determined by the choice of shell
manifold~\eqref{the.D.shells}.

The choice of $\mathcal{M}$ affects adapted coordinate system. The
line element corresponding to the metric $\gamma_{ij}$ in
\eqref{5,1} is
\begin{equation}
\gamma_{ij}\d x^i\d x^j =
\underbrace{\left(\d\vec{x}\right)^2}_{\displaystyle\mathbb{R}^{D-1}}
= \underbrace{\d r^{2}+r^{2}\d\Omega^2_{\beta}}_{
\displaystyle\mathbb{R}\times\mathbb{S}_{\beta}} +
\underbrace{\delta_{AB}\d x^A\d
x^B}_{\displaystyle\mathbb{R}^{D-2-\beta}}
,\label{adapted.gamma}\end{equation} where the coordinates are
labelled as follows
\begin{equation}
x^1 = r,\; \left\{x^{\Gamma} =
\theta_{\Gamma-1}\right\}^{\beta+1}_{\Gamma = 2},\; \theta_{\beta}
= \varphi,\; \left\{x^A = z_{A-\beta-1}\right\}^{D-1}_{A=\beta+2}
.\label{adapted.x}\end{equation} $\d\Omega^2_{\beta}$ denotes
volume element of a $\beta-$dimensional sphere in
equation~\eqref{adapted.gamma}. It can be written as
\begin{eqnarray*}
\d\Omega^2_{\beta} &=& \sum^{\beta-1}_{\Gamma=1}\left[
\prod^{\Gamma}_{\Upsilon=1}\sin^2\left(\theta_{\Upsilon}\right)\right]
\left(\d\theta_{\Gamma+1}\right)^2+\d\theta^2_{1}
.\end{eqnarray*} One can easily find that
\begin{equation}
\sqrt{|\gamma|} = r^{\beta}f(\theta),\; f_{,r} = 0
\label{gamma.determinant.r}\end{equation} holds in the adapted
coordinate system~\eqref{adapted.x}.

We assume the space--time has following symmetries -- static,
spherical symmetry of $\mathbb{S}_{\beta}$ and translational
symmetry along the $\mathbb{R}^{D-2-\beta}$ factor. These
symmetries correspond to following set of Killing vector fields,
expressed in the Cartesian coordinate system,
$$\frac{\partial\; }{\partial x^0};\;
x^i\frac{\partial\; }{\partial x^j} - x^j\frac{\partial\;
}{\partial x^i},\; 1\leq i, j\leq\beta+1;\; \frac{\partial\;
}{\partial x^A},\, A\in\{\beta+2,\ldots, D-1\}.$$

Now, we shall turn our attention to solving the differential
equation in~\eqref{5,51}.

The above discussed symmetries impose that the metric function $G$
depends only on the distance coordinate $r,$ i.e. $G~=~G(r),$ as
can be verified by explicit analysis of the Killing equations.

If we write the Laplacian acting on $G$ in the form presented
in~\eqref{Laplace.flat.general} and use that $G~=~G(r)$ together
with~\eqref{gamma.determinant.r}, we end up with
\begin{equation}
\Delta G(r) =
\frac{1}{r^\beta}\left(r^{\beta}G(r)_{,r}\right)_{,r},
\label{Laplace.flat.r.only}\end{equation} which in turn
implies~\eqref{5,51} takes the form
\begin{equation}
\frac{1}{r^\beta}\left(r^{\beta}G(r)_{,r}\right)_{,r}~=~-~\frac{D-3}{D-2}Kc^{2}G^{\frac{2}{D-3}}\rho_{\textup{m}},\;
\beta\in\{0,1,2,...\} ,\label{Laplace.Einstein}\end{equation}
where the mass density is given in
Table~\ref{shells.and.symmetry.D}.

\begin{table}[h]
\caption{The mass density of the shells
$\mathcal{M}$~\eqref{the.D.shells} in arbitrary $D~\geq~4$}
\begin{center}
{
\def\arraystretch{1.5}\begin{tabular}{|l|c|l|l|}\hline
Symmetry & $\beta$ & Number of shells & Mass density
$\rho_{\textup{m}}/\mu$
\\\hline\hline
(Hyper)Spherical & $D-2$ & 1 & $\delta(r-r_0)$
\\\hline
[Unnamed] & $D-2>\beta>1$ & 1 & $\delta(r-r_0)$
\\\hline
Cylindrical & $1$ & 1 & $\delta(r-r_0)$
\\\hline
Planar & $0$ & 2 (parallel) & $\delta(r-r_0)+\delta(r+r_0)$
\\\hline\end{tabular}
}
\end{center}
\label{shells.and.symmetry.D}\end{table}

Let us note the matter distribution can be generalised to a
perfect fluid distributed not only on a thin shell, $D~=~4$
spherical symmetry is examined e.g. in~\cite{Varela2,Varela3}.
Hyperspherical Majumdar-Papapetrou solutions containing thin
shells in general number of dimensions (the hypersphericity
implies $D=2+\beta$) are studied in~\cite{sijie} in a more general
setting. It is also possible to study thick (spherical)
shells~\cite{lemos}.

\bigskip
General form of solution to~\eqref{Laplace.Einstein} for
$|r|~\neq~r_0$ is
\begin{equation}
G(r) = C_1 + C_0\psi(r),\; \beta = 1: \psi = \ln r,\; \beta\neq 1:
\psi = \frac{|r|^{1-\beta}}{1-\beta}
,\label{G.general.beta.solution}\end{equation} where the absolute
value is added to include the planar symmetry case as will be
discussed later on.

Of course, interior and exterior values of both the integration
constants $C_1$ and $C_0$ are different in general. They are
determined by matching the interior and exterior (metric)
solutions which is done in the following.

We impose continuity of the metric, i.e. the function $G(r),$
across the shell. This condition can be expressed as
\begin{equation}
0 = \left.G\right|^{\rm ext}_{\rm int} = \lim_{\epsilon\rightarrow
0+}\left[ G(r_0+\epsilon) - G(r_0-\epsilon)\right] =
\left.C_1\right|^{\rm ext}_{\rm int} + \left.C_0\right|^{\rm
ext}_{\rm int}\psi(r_0) .\label{G.continuity}\end{equation} We
have used that $|r|~>~r_0$ correspond to exterior and $|r|~<~r_0$
correspond to interior.

Let us note that the constant $C_0$ is related to the total mass
of the shell. Indeed, we may integrate the
equation~\eqref{Laplace.Einstein} multiplied by $r^{\beta}$ across
the shell. The left hand side yields
$$\lim_{\epsilon\rightarrow 0+}\int^{r_0+\epsilon}_{r_0-\epsilon}
\left(r^{\beta}G(r)_{,r}\right)_{,r}\d r =
\left[r^{\beta}G(r)_{,r}\right]^{\rm ext}_{\rm int} =
\left.C_0\right|^{\rm ext}_{\rm int}.$$ The integrated right hand
side yields
$$\lim_{\epsilon\rightarrow 0+}\int^{r_0+\epsilon}_{r_0-\epsilon}
\frac{D-3}{D-2}Kc^{2}r^{\beta}G^{\frac{2}{D-3}}\rho_{\textup{m}}\d
r = \frac{D-3}{D-2}Kc^{2}r^{\beta}_0G(r_0)^{\frac{2}{D-3}}\mu,$$
the continuity of metric function $G$ across the shell was used.

Comparing both sides of the integrated equation leads to
\begin{equation}
\left.C_0\right|^{\rm ext}_{\rm int} =
-\frac{D-3}{D-2}Kc^{2}r^{\beta}_0G(r_0)^{\frac{2}{D-3}}\mu
.\label{C0.jump}\end{equation} The
equations~\eqref{G.general.beta.solution},~\eqref{G.continuity}
and~\eqref{C0.jump} allow us to write down the function $G$ in a
single formula using Heaviside step function $H$ as
\begin{equation}
G(r) = \underbrace{C_{1|\rm int} - \left.C_0\right|^{\rm ext}_{\rm
int}[ \psi(r_0)}_{\displaystyle C_{(r)1}}
\underbrace{-\psi(r)]H(r-r_0)+C_{0|\rm int}\psi(r)}_{\displaystyle
C_{(r)0}\,\psi(r)} ,\label{G.metric.function}\end{equation} where
the term $\left.C_0\right|^{\rm ext}_{\rm int}$ has already been
determined and it is given in~\eqref{C0.jump}.

The Heaviside step function $H$ satisfies three following
conditions
$$H(x>0) = 1,\; H(x<0) = 0,\; \frac{{\rm d}H(x)}{{\rm d}x} = \delta(x).$$

\medskip
In~\eqref{G.metric.function}, we have introduced another step
functions $C_{(r)i}$ that may change across the shell but are
constant otherwise, i.e.
$$C_{(r)i} = C_{i|{\rm int.}} + \left[
C_{i|{\rm ext.}} - C_{i|{\rm int.}}\right]H(r-r_0),\; C_{(r>r_0)i}
= C_{i|{\rm ext.}},\; C_{(r<r_0)i} = C_{i|{\rm int.}}.$$ These
functions naturally generalise the constants $C_1$ and $C_0.$
Because of their quite special dependence on $r,$ we write the $r$
as a lower case index.

With the Einstein equations coupling constant given by
$K~=~\frac{D-2}{D-3}\frac{4\pi\kappa}{c^{4}},$ see discussion in
footnote~\ref{coupling}, the equation~\eqref{G.metric.function}
becomes
\begin{eqnarray*}
G(r) &=& C_{1|\rm int} + C_{0|\rm int}\psi(r) +
\frac{4\pi\kappa\mu}{c^{2}}r^{\beta}_0G(r_0)^{\frac{2}{D-3}}\left[
\psi(r_0) - \psi(r)\right]H(r-r_0)
\\
&=& C_{(r)1} + C_{(r)0}\psi(r) .\end{eqnarray*}

Of course, the above result~\eqref{G.metric.function} holds for
one shell only and it followed from examination of the
differential equation in~\eqref{5,51}.

Let us consider the $\beta~=~0$ case. This time, $r$ can be
negative as well and we consider two shells located at
$r~=~\pm~r_0.$ We assume the space--time possesses a mirror
symmetry at $r~=~0,$ i.e. $G(-r)~=~G(r).$ Thus we consider $G$
depending on $r$ through $|r|$ only. A careful examination reveals
that the relation~\eqref{G.metric.function} still applies if we
replace $r$ by $|r|.$

\bigskip
We can summarize the results obtained thus far as follows
\begin{itemize}
\item
The metric is
\begin{equation}
\d s^2 = G^{-2}(c\d t)^2 - G^{2/(D-3)}\left[ \d
r^{2}+r^{2}\d\Omega^2_{\beta} + \delta_{AB}\d x^A\d x^B\right]
,\label{mp.shells.beta.general}\end{equation} where $G~=~G(r)$ is
given in~\eqref{G.metric.function}. In case of $\beta~=~0$ replace
all $r$ by $|r|.$

\item
The electrostatic potential $\phi_{\textup{e}}(r)$ is related to
$G(r)$ according to~\eqref{5,51}. The space--time dependence of
the electrostatic potential indicates that $F_{01}~=~-~F_{10}$ are
the only non--zero components of the electromagnetic field
strength tensor $F_{\alpha\beta}.$ These components correspond to
electrostatic field along the axis $x^1~=~r.$ Notice there is no
translational symmetry along that axis.

\item
The mass--charge balance condition~\eqref{5,6} holds.
\end{itemize}

Let us note that the number of shells, as long as each shell in
the set has the same symmetry, can be generalis/ed to higher values
in an obvious manner if centers of the spherical shells,
respectively axis of the cylindrical shells, coincide and all the
plane shells are parallel.

\section{Some examples with attentions to $D~=~4$}\label{shells.D.4.5}
We shall impose additional conditions on the general solution
derived in the preceding section.
\begin{enumerate}
\item
We assume the potential $\phi_{\textup{e}}$ is constant in the
interior (solution) which corresponds well with the
non--relativistic results obtained by use of Gauss electrostatic
law.
\begin{equation}
\left.\phi_{\textup{e}}\right|_{\rm int} = {\rm
const.}\Leftrightarrow \left.G_{,r}\right|_{\rm int} =
0\Leftrightarrow C_{(r<r_0)0} = 0\Rightarrow G(|r|\leq r_0) =
C_{(r<r_0)1} ,\label{potential.inside}\end{equation} where the
"$=$" (as a subcase of "$\leq$") in the last relation holds due to
continuity of the metric across the shell.

We see that the condition implies $G(|r|<r_0)$ is constant, hence
the interior region of space--time is flat and the metric can be
transformed into the Minkowski metric by appropriate rescaling of
the time and radial coordinates.

Thus it suffices to consider the exterior solution only.

\item
If $\beta~\geq~2,$ then $G(r~\rightarrow~\infty)$ does not diverge
as is clear from~\eqref{G.general.beta.solution}. We naively
impose asymptotical flatness condition by requiring
\begin{equation}
\beta\geq 2:\; \lim_{r\rightarrow\infty}G(r) = 1\Rightarrow
C_{(r>r_0)1} = 1 .\label{asymptotical.flatness}\end{equation} With
the general solution the solution for both $\psi(r)$ and $G(r),$
in equations~\eqref{G.general.beta.solution}
and~\eqref{G.metric.function}, and subsequent fixing of the
coupling constant $K$ we obtain
$$\beta\geq 2:\; 1 = G(r\rightarrow\infty) = C_{(r>r_0)1} = C_{(r<r_0)1}
- \frac{4\pi\kappa\mu}{c^{2}}r^{\beta}_0G(r_0)^{\frac{2}{D-3}}
\frac{1}{[\beta-1]r^{\beta-1}_0}.$$

The naive asymptotical flatness condition
$G(r\rightarrow\infty)~=~1$ cannot be satisfied for $\beta~\leq~1$
unless the exterior space--time region is flat. Indeed,
$\psi(r\rightarrow\infty)$ diverges and hence
$\left|G(r\rightarrow\infty)\right|$ is finite only if
$C_{(|r|>r_0)0}~=~0.$
\end{enumerate}

\medskip
The condition~\eqref{asymptotical.flatness}, applied in case of $\beta~\geq~2,$ fixes the value of $C_{(r>r_0)1}.$
In case of all admissible values of the parameter $\beta,$ we can suitably rescale the coordinates so that $C_1,$ either $C_{(r>r_0)1}$ or $C_{(r<r_0)1},$ is fixed again to $1~=~C_1$ the same value as implied on the $C_{(r>r_0)1}$ by the condition~\eqref{asymptotical.flatness}.

If $D-2-\beta~>~0,$ then there is at least one coordinate $z.$ The
total mass $M,$ as a volume integral of the mass density
$\rho_{\textup{m}},$ is infinite. For that reason, we introduce
mass per unit "volume" of the $z-$coordinates, that is finite, and
we denote it by $M_{D-2-\beta}.$ ($M_0~\equiv~M$ is the total mass
and it enters the metric function $G$ in the hyper--spherical
case, no $z-$coordinates.)

Then, the {\it exterior} metric function $G$ can be written using
the above introduced mass parameter $M_{D-2-\beta}$ as
$$G = C_{(r>r_0)1} + kM_{D-2-\beta}\psi(r),$$
where $k$ is constant determined by $\kappa,$ $c,$ $D,$ $\beta$ and a numerical factor.

Now, we shall perform the rescaling of spatial coordinates $r$ and
$z,$ or let us say a (spatial) length scale $L.$
\begin{eqnarray*}
L\rightarrow L' = C^{-1/(D-3)}_1L \Rightarrow
M_{D-\beta-2}\rightarrow M_{D-\beta-2}' &=&
C^{(D-\beta-2)/(D-3)}_1M_{D-\beta-2}
,\\
G\rightarrow G' &=& C_1G_1,
\end{eqnarray*}
where we have used the total mass $M$ is unaffected by the
rescaling and we have introduced abbreviations $G_1~=~G(C_1=1)$
and $C_1~=~C_{(r>r_0)1}$ for the sake of brevity.

Consequently, the coordinate expression for space--time metric
changes as
\begin{eqnarray*}
\d s^2 &=& G^{-2}(c\d t)^2 - G^{2/(D-3)}\left(\d\vec{x}\right)^2
\\
&\downarrow&
\\
\left(\d s'\right)^2 &=& C^{-2}_1G^{-2}_1(c\d t)^2 -
C^{2/(D-3)}_1G^{2/(D-3)}_1C^{-2/(D-3)}_1\left(\d\vec{x}\,'\right)^2
.\end{eqnarray*} The $C_1$ factors at the purely spatial
components cancel out so that it suffices to rescale the time
coordinate $t.$ Thus the total transformation of coordinates that
sets the integration constant $C_1$ from the exterior metric to
$1$ is
\begin{equation}
t\rightarrow t' = C_1t,\; r\rightarrow r' = C^{-1/(D-3)}_1r,\;
z\rightarrow z' = C^{-1/(D-3)}_1z
.\label{coordinate.transform.C.1}\end{equation} We drop the primes
in the following.

Thus we have transformed away the constant $C_{(|r|>r_0)1}.$
Similar rescaling can be performed for $|r|~<~r_0$ to "remove" $C_{(|r|<r_0)1},$ this is what we do in case of $\beta\{0,~1\}.$

By the first condition and continuity of the metric across the shell,~\eqref{potential.inside}, it follows that the interior and exterior $C_1$'s are related by linear transformation.
Hence the rescaling that sets either the exterior or the interior $C_1$ equal one removes undeterminacy of the remaining $C_1.$

\bigskip
{\it The conditions~\eqref{potential.inside} and
either~\eqref{asymptotical.flatness} (in case of $\beta~\geq~2$)
or the rescaling of coordinates~\eqref{coordinate.transform.C.1}
(in the interior in case of $\beta~\in~\{0,1\}$) are implicitly imposed on all the
sub--cases examined in the following. }


\bigskip
There are three possible distinct values of $\beta$ in case of
$D~=~4$ -- the three symmetries summarized in
Table~\ref{flat.and.symmetry}.

\begin{table}
\caption{Symmetries and adapted coordinate systems in $D~=~4$}
\begin{center}
{
\def\arraystretch{1.5}\begin{tabular}{|l|c|c|l|}\hline
Symmetry & Shell $\mathcal{M}$ & Coord. system &
$(\d\vec{x})^2~=~\gamma_{ij}\d x^i\d x^j$
\\\hline\hline
Spherical & $\mathbb{S}_{2}$ & $(ct,r,\theta,\varphi)$ & $\d
r^2+r^{2}\d\theta^2+r^{2}\sin^{2}(\theta)\d\varphi^2$
\\\hline
Cylindrical & $\mathbb{R}\times\mathbb{S}$ & $(ct,r,\varphi,z)$ &
$\d r^2+r^{2}\d\varphi^2+\d z^2$
\\\hline
Planar & $\mathbb{R}^{2}$ & $(ct,r,z_{1},z_{2})$ & $\d r^2+\d
z_{1}^2+\d z_{2}^2$
\\\hline\end{tabular}
}
\end{center}
\label{flat.and.symmetry}\end{table}


\subsection{$D~=~4$ spherical shell ($\beta~=~2$)}\label{D.4.beta.2}
The solution~\eqref{G.metric.function}, subject to
both~\eqref{potential.inside} and~\eqref{asymptotical.flatness},
combined with expression for total mass~\eqref{5,61}
$M=4\pi r_{0}^{2}G^2(r_0)\mu$ 
reduces to
\begin{equation*}
\mathrm{d} s^2=\frac{1}{G^2}(c\mathrm{d} t)^2-G^2\left[\mathrm{d}
r^2+r^{2}\mathrm{d} \theta^2+r^2\sin^2(\theta)\mathrm{d}
\varphi^2\right],
\end{equation*}
\begin{equation}
G(r\geq r_{0})=1+\frac{M\kappa}{c^{2}r},\; G(r\leq
r_{0})=1+\frac{M\kappa}{c^{2}r_{0}},\quad
\frac{M\kappa}{c^2} = \frac{1}{2}R_{s}
\label{16}
\end{equation}
The exterior solution describes extremal Reissner-Nordstr\"{o}m
space--time and the constant $\frac{2M\kappa}{c^2}$ in~\eqref{16}
can be related to the Schwarzschild radius $R_{s}$ as given above.

\subsection{$D~=~4$ cylindrical shell ($\beta~=~1$)}\label{D.4.beta.1}
Cylindrical space-times were studied e.g.
in~\cite{stephani,podolsky}.

Total mass $M$ and charge $Q$ are infinite. Therefore, we
introduce corresponding quantities related to a unit length (along
the axis) of the cylinder
$$M_1 = M/{\rm length} = 2\pi r_0G^2(r_0)\mu,$$
similarly for the charge. Then we can rewrite the metric function
$G$ in terms of $M_1$ as
\begin{equation}
G(r) = 1 -
\frac{2M_1\kappa}{c^2}\ln\left(\frac{r}{r_0}\right)H(r-r_0)
,\label{24}\end{equation}

where line element, in the rescaled coordinates is
\begin{equation*}
\mathrm{d} s^2= \frac{1}{G^2}(c\mathrm{d}t)^2 - G^2\left[
\mathrm{d} r^2+r^2\mathrm{d}\varphi^2+\mathrm{d} z^2\right].
\end{equation*}

\subsection{$D~=~4$ two parallel plane shells ($\beta~=~0$)}\label{D.4.beta.0}
We obtain $G(|r|~\leq~r_0)~=~C_{(|r|<r_0)1}$ as in the
Section~\ref{D.4.beta.1}. The constant $C_{(|r|<r_0)1}$ is set to
$1$ by the coordinate
transformation~\eqref{coordinate.transform.C.1}.
\begin{equation}
G(r) = G(|r|) = 1 + \frac{4\pi\kappa G^2(r_0)\mu}{c^{2}}\left[
r_0-|r|\right]H(|r|-r_0) .\label{31,321}\end{equation} Hence the
resulting exterior line element is
\begin{equation*}
\mathrm{d} s^2 =\frac{1}{G^2}(c\mathrm{d}t)^2 - G^2\left[
\mathrm{d}r^2+\mathrm{d}z_{1}^2+\mathrm{d}z_{2}^2\right]
\end{equation*}
with $G$ given by~\eqref{31,321}.

\subsection{$D~=~5$ hyperspherical shell ($\beta~=~3$)}
Solution of the $D~=~5$ hyperspherical shell with $G(r)$ (for
$D=5$ and $\beta=3$) subject to both~\eqref{potential.inside}
and~\eqref{asymptotical.flatness} is
\begin{equation}
G(r) = 1 +
\frac{4\pi\kappa\mu}{c^2}r^3_0G(r_{0})\left(\frac{1}{2r^2_0} +
\left[\frac{1}{2r^{2}}-\frac{1}{2r_{0}^{2}}\right]H(|r|-r_0)\right)
.\label{32}\end{equation} Total mass of the hyperspherical shell
is given by $M=2\pi^{2} r_{0}^{3}G(r_0)\mu.$ The line element
therefore can be written as
\begin{equation*}
\mathrm{d} s^2=\frac{1}{G^2}(c\mathrm{d} t)^2-G\left[\mathrm{d}
r^2+r^{2}\mathrm{d}
\theta_{1}^2+r^2\sin^2(\theta_{1})\mathrm{d}\theta_{2}^{2}
+r^2\sin^2(\theta_{1})\sin^2(\theta_{2})\mathrm{d}\varphi^2\right],
\end{equation*}
\begin{equation}
G(r\geq r_{0})=1+\frac{M\kappa}{\pi c^{2}r^{2}},\; G(r\leq
r_{0})=1+\frac{M\kappa}{\pi c^{2}r_{0}^{2}}.
\label{33}\end{equation}
The exterior solution describes extremal Reissner-Nordstr\"{o}m space--time in $D=5.$

\section{$D~=~4$ non--relativistic approximation}\label{non.relativistic}
Non--relativistic, i.e. classical, mechanics of thin shells will
be treated in this section.

Let us consider two small pieces of the shell(s) with surface
areas $\mathrm{d}S,$ $\mathrm{d}S',$ their mutual distance being
$a.$ Charge and mass surface densities of the shell(s) are $\mu$
and $\sigma.$

The pieces mutually interact by means of attractive gravitational
and repulsive electrostatic forces.

The total force piece $\mathrm{d}S$ acts on the piece
$\mathrm{d}S'$ is
$$\d\vec{F} =
- \underbrace{\kappa\frac{\mu^{2}}{a^{2}}\d S\d S'\vec{n}}_{\rm
gravitational} +
\underbrace{\frac{\sigma^{2}}{4\pi\varepsilon_{0}a^{2}} \d S\d
S'\vec{n}}_{\rm electrostatic} =
\left[-\kappa\mu^{2}+\frac{\sigma^{2}}{4\pi\varepsilon_{0}}\right]
\frac{1}{a^2}\d S\d S'\vec{n},$$ where $\vec{n}$ is unit length
vector pointing from $\mathrm{d}S$ to $\mathrm{d}S'.$

If we demand the shell is static, it follows $\d\vec{F}~=~0,$
hence the balance condition
$\frac{\sigma^{2}}{4\pi\varepsilon_{0}}=\kappa\mu^{2}.$ This
condition has the same form also for general relativity in
four--dimensional space-time.

\medskip
Let us turn our attention to equation~\eqref{5,51} which is an
example of Poisson equation.

Every $D~=~4$ solution of the shells considered in this work can
be written in the form $G=1-\frac{\mu f(r)}{c^{2}}$. In the
non-relativistic limit, one has $\mu f(r)\ll c^{2}$ (hence
$G~\doteq~1$), this may be due to weak gravitational source and /
or sufficiently large distance from the source, i.e. low $f(r).$

If the non--constant term in $G$ is small, then the Poisson
equation simplifies to
$$-\Delta[\mu f(r)] = \Delta Gc^2 = -~\frac{1}{2}Kc^{4}G^2\rho_{\textup{m}}\doteq -4\pi\kappa\rho_{\textup{m}}.$$
This equation is similar to well known Poisson equation for
non-relativistic gravitational field.

Let us introduce a new function
$\phi_{\textup{m}}~=~\mu~f(r)~+~K_{m}.$ With this definition, the
Poisson equation becomes
$$\Delta\phi_{\textup{m}}=4\pi\kappa\rho_{\textup{m}}.$$
This is precisely the classical mechanics Poisson equation for the
gravitational potential $\phi_{\textup{m}}$ that is determined up
to an additional constant $K_{m}.$

Concrete examples of gravitational potentials of massive shells
can be read off from equations \eqref{16}, \eqref{24} and
\eqref{31,321}. The potential is in case of a spherical shell
\begin{equation}
\phi_{\textup{m}}(r\geq r_{0})=-4\pi
\kappa\frac{r_{0}^{2}\mu}{r}+K_{m}, \phi_{\textup{m}}(r\leq
r_{0})=-4\pi \kappa r_{0}\mu+K_{m},\label{38}
\end{equation}
in case of a cylindrical shell
\begin{equation}
\phi_{\textup{m}}=4\pi \kappa
r_{0}\mu\ln\left(\frac{r}{r_{0}}\right)H(r-r_{0})+K_{m} \label{40}
\end{equation}
and in case of two planar shells
\begin{equation}
\phi_{\textup{m}}=4\pi \kappa\mu[|r|-r_{0}]H(|r|-r_{0})+K_{m}.
\label{42}
\end{equation}
This is in agreement with the results of classical mechanics.

A general relation between non-relativistic potential
$\phi_{\textup{m}}$ and the metric function $G$ is given by
\begin{equation}
\phi_{\textup{m}}=c^{2}(1-G)+K_{m}
.\label{potential.G}\end{equation} As a result, the metric
coefficients $g_{\alpha\beta}$ of Majumdar--Papapetrou form can be
easily obtained if we know the non-relativistic potential.

\section{Remark on generalisation to Kastor--Traschen}\label{Kastor.Traschen.lambda}
The Majumdar--Papapetrou space--time can be generalised to a
co--moving charged dust and a positive cosmological constant.

Adding a cosmological constant is a motivation to consider a
slightly generalised metric ansatz, when compared to~\eqref{5,1},
with $G$ allowed to depend on time.
\begin{equation}
\d s^2 = G^{-2}(c\,\d t)^2 - G^{2/(D-3)}\,\gamma_{ij}\d x^i\d
x^j,\; G = G(t, x^i) .\label{Kastor.Traschen.metric}\end{equation}
The above metric can also be expressed in an alternate form
$$\d s^2 = \bar{G}^{-2}(c\,\d\bar{t})^2
- \bar{G}^{2/(D-3)}{\rm e}^{2a(t)}\,\gamma_{ij}\d x^i\d x^j$$ that
generalises the $D~=~4$ metric ansatz of~\cite{metin2} and it is
related to~\eqref{Kastor.Traschen.metric} by following
transformations
$$\d\bar{t} = {\rm e}^{-(D-3)a(t)}\d t,\; \bar{G} = G{\rm e}^{-(D-3)a(t)}.$$

The difference $\Delta_{\mu\nu}~\equiv~G_{\mu\nu}~-~G^{\rm
MP}_{\mu\nu}$ of the Einstein tensor of the above
metric~\eqref{Kastor.Traschen.metric} and the $G^{\rm
MP}_{\mu\nu}$ corresponding to the Majumdar--Papapetrou metric is
\begin{equation}
\Delta_{00} = \frac{\delta_D\epsilon_D}{2}\dot{G}^2g_{00},\;
\Delta_{0i} = \delta_D\frac{G_{,0i}}{G},\; \Delta_{ij} =
\frac{\delta_D}{2}\left[2G\ddot{G} +
\epsilon_D\dot{G}^2\right]g_{ij}
,\label{Kastor.Traschen.Einstein}\end{equation} where dimension
dependent factors
$$\delta_D\equiv\frac{D-2}{D-3},\;\epsilon_D\equiv\frac{D-1}{D-3}$$
were introduced for the sake of brevity.

\medskip
The mixed component of the Einstein equations implies
\begin{equation}
G_{0i} = 0\Rightarrow G_{,0i} = 0\Rightarrow G = T(t) + R(x^i)
.\label{G.split}\end{equation}

The extra terms in $G_{00}$ and $G_{ij},$ when compared to the
case of $\Lambda~=~0,$ are proportional to metric and hence they
can be attributed to the cosmological constant but only for a suitable
choice of $G$ given by
\begin{equation}
G = \pm\sqrt{\frac{2\Lambda}{\delta_D\epsilon_D}}t + R(x^i) =
\pm\sqrt{\frac{2\Lambda}{(D-1)(D-2)}}(D-3)t + R(x^i)
.\label{G.Lambda}\end{equation} The metric function $G$ presented
in~\eqref{G.Lambda} ensures that the relations among
$\phi_{\textup{e}}(G),$ $\rho_{\textup{e}},$ $\rho_{\textup{m}}$
and $G$ (namely the equations~\eqref{5,51} and~\eqref{5,6}) have
formally the same form as in the case of $\Lambda~=~0,$ the only
difference being that constants of integration with respect to
spatial coordinates are functions of time in general.

It is immeadiate to see
that the Laplace operator (of the metric
$\gamma_{ij}$) acting on $G$ gives $\Delta~G~=~\Delta~R,$ i.e. $R$
is determined by the classical potential $\phi_{\textup{m}}$ in a
similar way as was the pure Majumdar--Papapetrou $G,$ see
equation~\eqref{potential.G}.

\section{Conclusion}\label{conclusion}
We have presented a systematic study of static higher dimensional
thin charged dust hypersurface shells of the form
$\mathbb{S}^{\beta}\times\mathbb{R}^{D-2-\beta}$ in arbitrary
$D~\geq~4$ space--time of the Majumdar--Papapetrou form. The
metric is determined by a single function denoted, in this work,
by $G.$

The specific form of the Majumdar--Papapetrou metric reduced the
Einstein--Maxwell equations to Laplace equation for the single
unknown metric function in the source--free regions. The
electrostatic potential $\phi_{\textup{e}}$ was determined by the
metric function $G$ and consistency of the equations implied
relation between mass and charge densities of the dust.

The Poisson--like relation between $G$ and mass density
$\rho_{\textup{m}}$ implied the metric function $G$ can be built
using a non--relativistic gravitational potential
$\phi_{\textup{m}},$ the relation is presented in
equation~\eqref{potential.G}.

\medskip
The obtained results can be further generalised beyond the already
addressed Kastor--Traschen space--time. Possible generalisation of
the results presented in this article to the case of $\gamma_{ij}$
or $\d\Omega^2,$ the former introduced in equation~\eqref{5,1},
being a maximally symmetric metric of arbitrary constant curvature
is currently under investigation. It is also possible to study
shells that are not hypersurfaces, i.e. $\dim\mathcal{M}~<~D~-~2,$
or generalise the results to modified theories of gravity.

\section*{Acknowledgements}
Both authors are grateful to both prof. Jan Novotn\'{y} and to the
Department of Physical Electronics for their ongoing support.

Financial support of M. \v{C}erm\'{a}k by the Academy of Sciences of the Czech Republic, Project No. AV0Z 20410507, is acknowledged.



\end{document}